\documentclass[12pt]{article}
\usepackage{amssymb}
\usepackage{amsthm}
\usepackage{amsmath}
\usepackage{amsfonts}
\usepackage{graphicx}
\usepackage{titlesec}

\titleformat*{\section}{\normalfont\bfseries}

\begin{document}
\title{Some comments on the structure of the best known networks sorting 16 elements}
\date{}
\author{Igor S. Sergeev\footnote{e-mail: isserg@gmail.com}}
\maketitle

\begin{abstract}
We propose an explanation of the structure of the best known
sorting networks for 16 elements with respect to the complexity
and to the depth due to Green and van Voorhis.
\end{abstract}

\section{Introduction}

Consider the standard sorting networks model. Such networks are
built of comparators ordering a pair of elements. All network
inputs and comparators have fan-out~1. Standard measures applied
to networks are complexity (the number of comparators) and depth
(the number of layers of independent comparators). A comprehensive
study of notions and theory of sorting networks can be found
in~\cite{knu07e}.

Efficient sorting networks for small number of inputs were
discovered until 1970s. Nowadays computers are intensively
involved both into the search of new optimal networks, and into
the verification of optimality of already known networks.

Among the known examples one can distinguish 16-element sorting
networks proposed by Green~\cite{gr69e} and van Voorhis about
1970, see~\cite{knu07e}. First, because 16 is a power of 2.
Second, these constructions are not simply reduced to combination
of sorting networks of a smaller size. Third, the networks were
discovered by human (without computer help, or with just a little
help). Seemingly, 16 is a maximal number to satisfy all above
conditions.

The Green network has complexity 60 and depth 10. Known
computational experiments confirm the optimality of the complexity
bound with almost hundred-percent probability. The van Voorhis
network has complexity 61 but smaller depth~9. Recently,
in~\cite{bc14e} this depth bound was proved to be tight.

An example of a depth-9 network is of additional interest due to
the corollary on existence of a Boolean circuit of the same depth
for majority functions of 15 or 16 variables. These functions are
important for implementation of fast integer multipliers. Though
Boolean circuits serve as more flexible and powerful tool for
computations, alternative ways to achieve depth 9 (almost
definitely, optimal) for these functions are probably not known.

The author is not familiar with any publication providing an
analysis of the structure of Green and van Voorhis networks. So,
he believes that a bit of explanation proposed here may be
suitable.

\section{Approximate sorting phase}

We follow a standard way of graphic representation of sorting
networks, where elements correspond to horizontal lines, and
comparators are depicted as bridges connecting two lines,
see~\cite{knu07e}. One can easily check (or consult
with~\cite{knu07e}) that the network shown on Pict.~1 generally
orders $2^n$ inputs into a Boolean cube poset (the picture
represents the case $n=3$).

\begin{center}
\begin{tabular}{l}
\includegraphics[scale = 0.5, bb= 0 0 250 255]{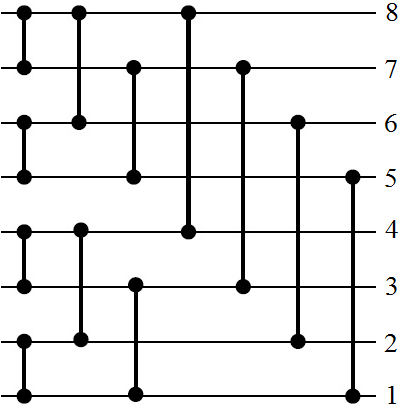}   %
\end{tabular}

Pict. 1. Approximate sorting network for 8 elements
\end{center}

Apparently, the said network was introduced by Green as an
efficient initial phase of sorting. A thorough analysis of such
approximate sorting networks can be found in~\cite{lp98e}.

So, efficient 16-element sorting networks start with the
approximate sorting subnetwork. Its depth is 4, and complexity is
32. The intermediate result is the ordering of elements as shown
on Pict.~2. Graph nodes are labeled with the numbers of lines of
the standard network representation, where the corresponding
elements locate. Maximum is on the top, and minimum is at the
bottom. From now on, we denote elements by their numbers according
to the picture. Also, Roman numerals are used to number layers of
the Boolean cube.

\begin{center}
\begin{tabular}{l}
\includegraphics[scale = 0.5, bb= 0 0 595 455]{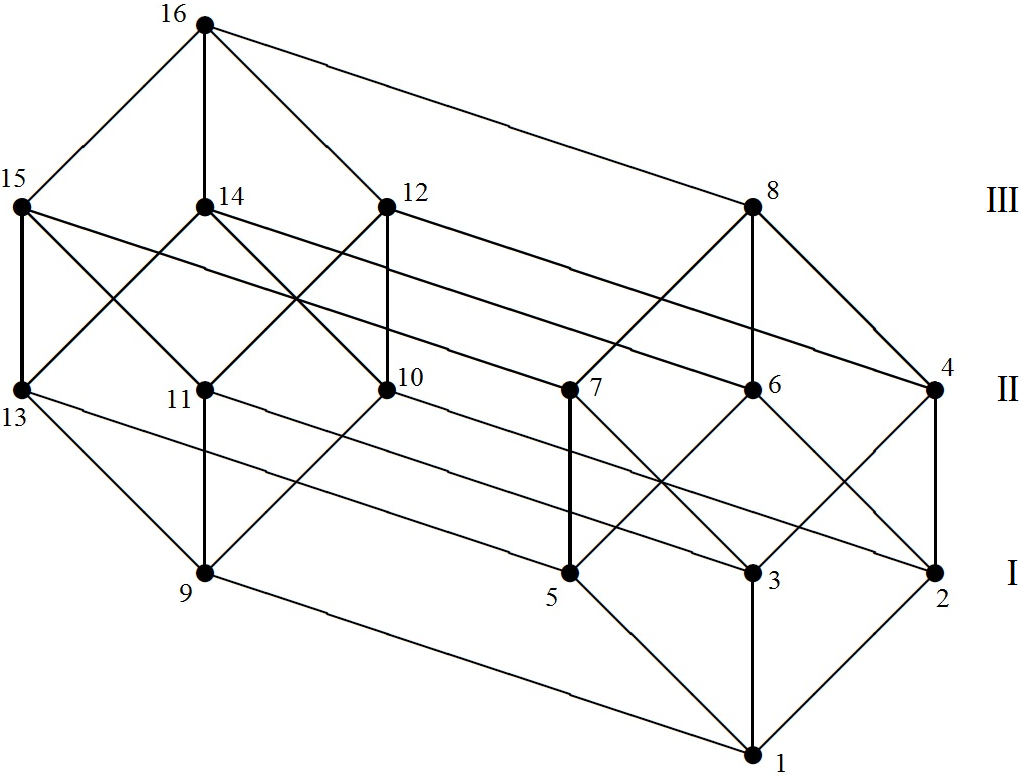}   %
\end{tabular}

Pict. 2. Elements' ordering after the approximate sorting phase
\end{center}

Let us make a few simple observations.

a) Evidently, the largest and the smallest elements are already
determined (16 and 1 correspondingly).

b) The second and the third largest elements are two maximal
elements of the layer III. By symmetry, the second and the third
smallest elements are two minimal elements of the layer I.

c) Six medial elements belong to an 8-element set $M$ containing
the middle layer of the cube, the smallest element of the layer
III, and the largest element of the layer I.

d) The fourth and the fifth largest elements are the third largest
element of the layer III and maximal element in $M$. A symmetric
statement holds for the lower part of the order.

These observations lead to the following strategy. Sort layer I,
layer III, set $M$, and compare the maximum of $M$ with the third
element of the layer III, and the minimum of $M$ with the second
element of the layer I.

Tetrads of layer-I and layer-III elements can be sorted by simple
depth-3 networks of 5 comparators, computing maximal and minimal
elements at the depth~2.

Green and van Voorhis networks differ only in the way they sort
the set $M$.

\section{Sorting of $M$. Low-complexity way}

The first level of the next stage (depth-5 comparators) is common
for both Green and van Voorhis methods. There are compared
elements in pairs (13, 4), (11, 6), (10, 7). Let (13', 4'), (11',
6'), (10', 7') denote ordered pairs.

The pairs are composed in the way to neighbor with all elements of
layers I and III. Thus, winners of the comparisons (that is,
elements 13', 11', 10') outmatch all elements of lower layers of
the cube, while loosers are outrun by all elements of upper
layers. Denote $8'=\min\{15, 14, 12, 8\}$ and $9'=\max\{9, 5, 3,
2\}$ (elements of $M$ prepared at the depth 6).

Green network then sorts tetrads $\{13', 11', 10', 8'\}$, $\{9',
7', 6', 4'\}$, and merges them afterwards.

\begin{center}
\begin{tabular}{l}
\includegraphics[scale = 0.5, bb= 0 0 180 360]{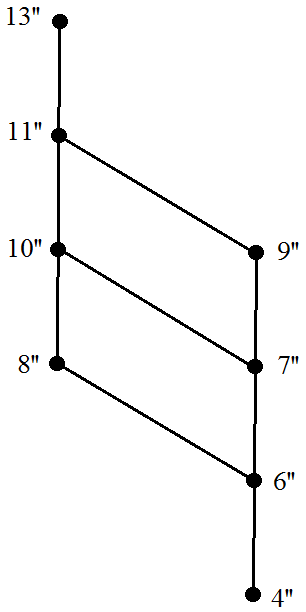}   %
\end{tabular}

Pict. 3. Ordering of elements in $M$ after sorting of tetrads
\end{center}

Let us denote elements of ordered tetrads by numbers with two
strokes. The partial order of elements after sorting of tetrads is
shown on Pict.~3. It is such, since any element of the first
tetrad beats some two elements of the second tetrad, and the third
largest element outmatches three elements of the second tetrad.
Conversely, any element in the second tetrad loses to some two
elements of the first tetrad.

Therefore, two largest and two smallest elements in $M$ are
already determined, and it remains to complete ordering of the
tetrad $\{10'', 9'', 8'', 7''\}$. This step can be done via 3
comparisons. For saving some depth it makes sense to compare
elements $9''$ and $8''$ first, since these elements are available
at the depth 8, while other two elements only at the depth 9.

\section{Sorting of $M$. Low-depth way}

Similarly, van Voorhis network also proceeds with computing
elements $13'$, $11'$, $10'$, $7'$, $6'$, $4'$ at the depth 5.
However, since elements $8', 9'$ are not ready yet, on the next
level there executed additional comparisons in pairs (13', 6'),
(11', 7'), (10', 4'). Let numbers with two strokes denote the
results of these comparisons. Then, after application of six
levels of comparators elements of $M$ are ordered in a poset shown
on Pict.~4.

\begin{center}
\begin{tabular}{l}
\includegraphics[scale = 0.5, bb= 0 0 372 186]{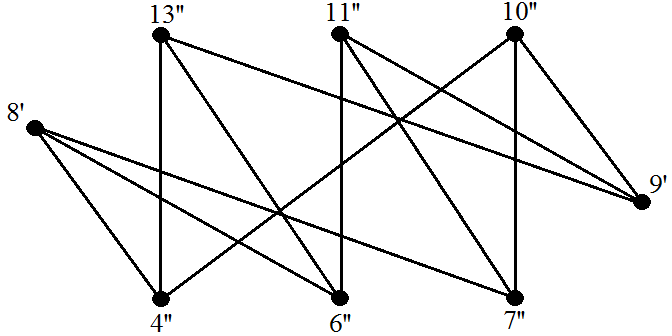}   %
\end{tabular}

Pict. 4. Ordering of elements in $M$ after preliminary comparisons
\end{center}

Easy to see that any element of the upper tetrad $\{8', 13'',
11'', 10''\}$ outmatches some three elements of the lower tetrad
$\{4'', 6'', 7'', 9'\}$, and vice versa: any element of the lower
tetrad looses to some three elements of the upper tetrad.

Hence, the largest three elements in $M$ belong to the upper
tetrad, the smallest three --- to the lower tetrad; minimum in the
upper tetrad and maximum in the lower tetrad constitute two medial
elements.

To finalize the network one has to sort these tetrads and
implement an extra comparison to order medial elements.

\section{Supplement}

Pict.~5, 6 show Green and van Voorhis networks. A vertical line
separates approximate sorting phase into the Boolean cube poset.
Subnetworks sorting layers I and III are labeled by 1 and 2,
respectively. Subnetworks sorting tetrads $\{13', 11', 10', 8'\}$
or $\{13'', 11'', 10'', 8'\}$ are labeled by 3, and subnetworks
sorting tetrads $\{9', 7', 6', 4'\}$ or $\{9', 7'', 6'', 4''\}$
are labeled by 4. The final merging subnetwork in the Green
network is labeled by 5.

\begin{center}
\begin{tabular}{l}
\includegraphics[scale = 0.4, bb= 0 0 945 497]{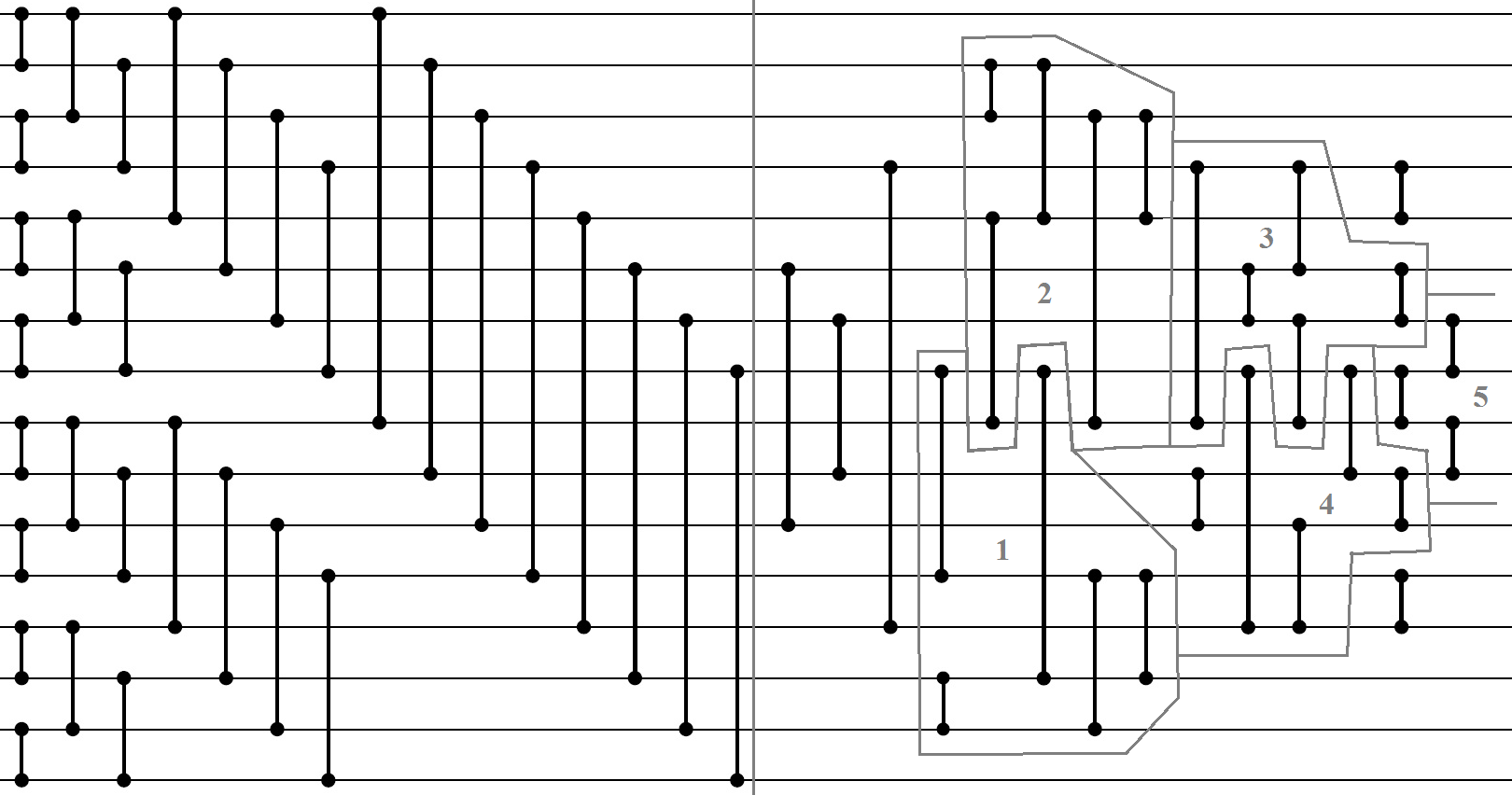}   %
\end{tabular}

Pict. 5. Green network
\end{center}

\begin{center}
\begin{tabular}{l}
\includegraphics[scale = 0.4, bb= 0 0 976 504]{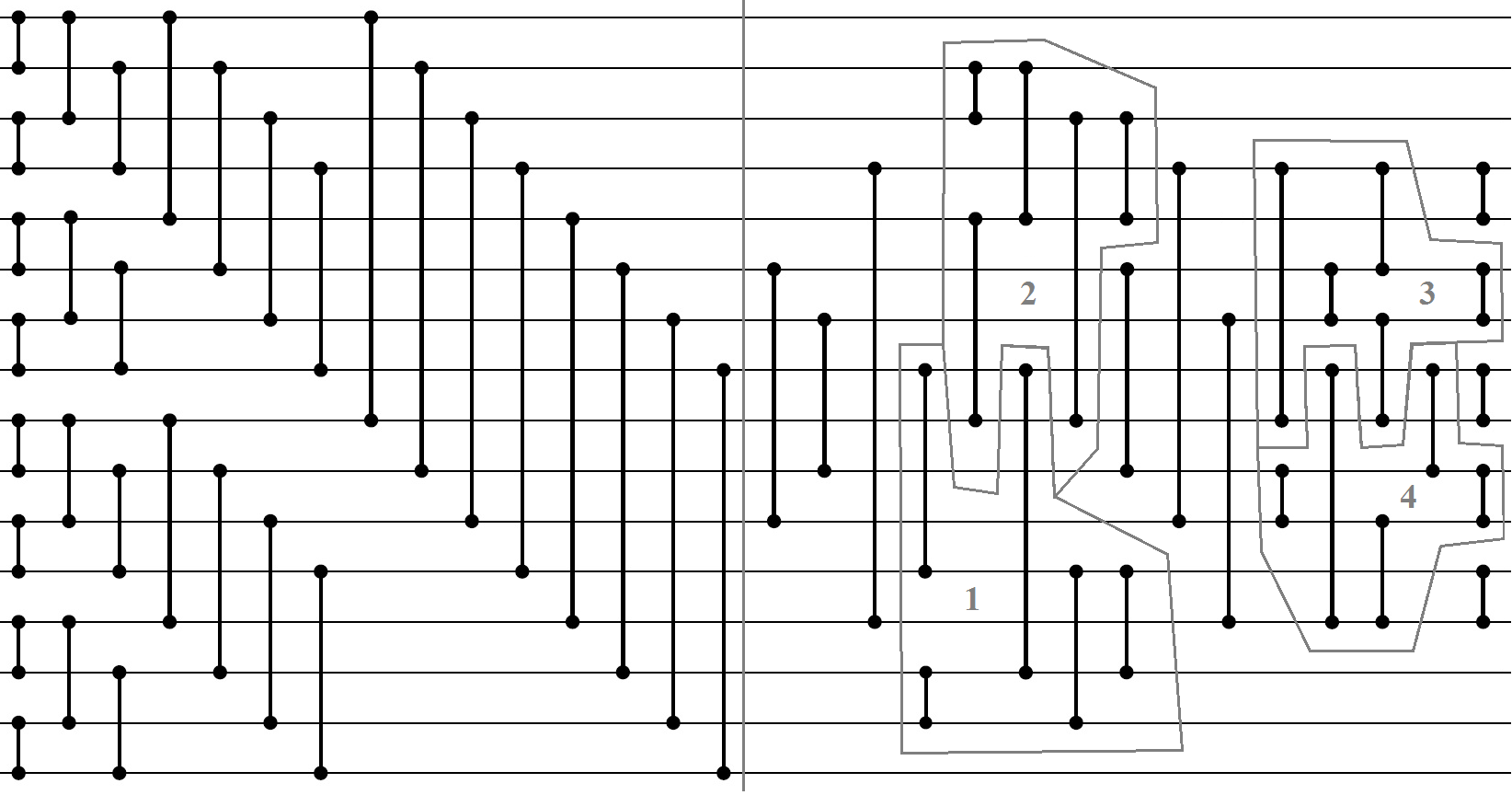}   %
\end{tabular}

Pict. 6. van Voorhis network
\end{center}

Papers~\cite{jui95e,cm02e,val13e} provide discovered with the use
of computer additional examples of networks of depth 10 and
complexity 60 with different structure. In particular, these
networks contain substantially reduced phase of the Boolean cube
approximate sorting.

The present work is partially supported by RFBR grant no.
17-01-00485a.

\end{document}


\title{Пояснение принципа работы наилучших схем сортировки 16 элементов}
\date{}
\author{И. С. Сергеев\footnote{эл. почта: isserg@gmail.com}}
\maketitle

\begin{abstract}
Предлагается пояснение принципа работы наилучших по сложности и по
глубине схем сортировки 16 элементов Грина и ван Вориса.
\end{abstract}

\section{Введение}

Мы рассматриваем стандартные схемы сортировки. Они составляются из
компараторов, упорядочивающих пару элементов. Ветвления входов
схемы и выходов компараторов запрещены. Основными параметрами
схемы являются сложность (число компараторов) и глубина (число
слоев из параллельно работающих компараторов). Подробное изложение
понятий и теории схем сортировки см. в~\cite{knu07}.

Эффективные схемы для сортировки небольшого числа элементов были
найдены еще к 1970-м гг. Сегодня как к построению оптимальных
схем, так и к доказательству оптимальности уже найденных активно
привлекаются ЭВМ.

Среди известных примеров эффективных схем выделяются схемы
сортировки 16 элементов, найденные Грином~\cite{gr69} и ван
Ворисом около 1970 г., см.~\cite{knu07}. Во-первых, потому что 16
--- степень двойки. Во-вторых, эти конструкции прямо не сводятся к
манипуляциям со схемами сортировки меньшей размерности. В-третьих,
они найдены человеком (без помощи ЭВМ или с минимальной помощью).
Похоже, что 16 --- это максимальное число, для которого верно все
перечисленное.

Схема Грина имеет сложность 60 и глубину 10. Известные результаты
вычислительных экспериментов показывают почти со стопроцентной
вероятностью оптимальность оценки сложности. Схема ван Вориса
имеет сложность 61, но глубину 9. Недавно в~\cite{bc14} доказано,
что схема оптимальна по глубине.

Пример схемы глубины 9 интересен еще и в связи со следствием о
существовании булевой схемы такой же глубины для функций
голосования 15 или 16 переменных --- эти функции в свою очередь
важны для построения быстрых умножителей чисел. Хотя булевы схемы
представляют собой гораздо более гибкий и мощный инструмент для
вычислений, другие способы достижения оценки 9 (почти наверняка,
неулучшаемой) для этих функций, видимо, неизвестны.

Автору не встречались публикации с анализом структуры, а заодно и
корректности работы схем Грина и ван Вориса. Поэтому он надеется,
что приведенные в настоящей работе пояснения могут быть полезны
для понимания природы оптимальных схем сортировки.

\section{Приближенная сортировка}

Мы используем стандартный способ графического представления схем
сортировки, в котором элементы соответствуют горизонтальным
линиям, а компараторы изображаются перемычками, соединяющими две
линии, см.~\cite{knu07}. Несложно проверить (или посмотреть
в~\cite{knu07}), что изображенная на рис. 1 схема в общем случае
упорядочивает $2^n$ входов в топологии булева куба (на рисунке
показан случай $n=3$).

\begin{center}
\begin{tabular}{l}
\includegraphics[scale = 0.5, bb= 0 0 250 255]{partsort.png}   %
\end{tabular}

Рис. 1. Схема приближенной сортировки 8 элементов
\end{center}

Указанная схема, вероятно, введена Грином как эффективная
начальная стадия сортировки. Подробный анализ таких схем
приближенной сортировки выполнен в~\cite{lp98}.

Итак, эффективные схемы сортировки 16 элементов начинаются с
выполнения приближенной сортировки --- ее глубина равна 4, а
сложность 32. После этого элементы упорядочены, как показано на
рис. 2. Вершины пронумерованы номерами линий, на которых
оказываются элементы при графическом изображении схемы.
Максимальный элемент сверху, минимальный снизу. Далее элементы
будем обозначать номерами согласно рисунку. Римскими цифрами
занумерованы слои булева куба.

\begin{center}
\begin{tabular}{l}
\includegraphics[scale = 0.5, bb= 0 0 595 455]{boolean.png}   %
\end{tabular}

Рис. 2. Порядок элементов после приближенной сортировки
\end{center}

Сделаем несколько простых наблюдений.

а) Очевидно, максимальный и минимальный элемент уже определены (16
и 1 соответственно).

б) Вторым и третьим по величине являются два наибольших элемента
слоя III. Симметрично, вторым и третьим с конца являются два
наименьших элемента слоя I.

в) Шесть средних по величине элементов в наборе принадлежат
множеству $M$ из 8 элементов, включающему средний слой куба,
минимальный элемент слоя III и максимальный элемент слоя I.

г) Четвертым и пятым по величине являются третий по величине
элемент слоя III и максимальный элемент множества $M$.
Симметричное утверждение справедливо для нижней части порядка.

Из этих наблюдений вытекает следующая стратегия. Выполнить
сортировку слоя I, слоя III, множества $M$ и сравнить максимум $M$
с 3-м элементом слоя III, а минимум $M$ со вторым элементом слоя
I.

Четверки элементов слоев I и III сортируются простой схемой
сложности 5 и глубины 3, причем максимальные и минимальные
элементы определяются с глубиной 2.

Схемы Грина и ван Вориса отличаются только способом сортировки
множества $M$.

\section{Cортировка множества $M$. Способ Грина}

Первый уровень следующей фазы сортировки (компараторы на глубине
5) у Грина и ван Вориса совпадает. На нем выполняются сравнения в
парах (13, 4), (11, 6), (10, 7). Элементы упорядоченных пар
обозначим (13', 4'), (11', 6'), (10', 7').

Пары сформированы с таким умыслом, чтобы иметь соседями все
элементы слоев I и III. Поэтому победители сравнений (элементы
13', 11', 10') превосходят все элементы нижних слоев куба, а
проигравшие уступают всем элементам более верхних слоев.

Обозначим $8'=\min\{15, 14, 12, 8\}$ и $9'=\max\{9, 5, 3, 2\}$
(элементы множества $M$, которые вычисляются на глубине 6).

В методе Грина выполняется сортировка в четверках $\{13', 11',
10', 8'\}$ и $\{9', 7', 6', 4'\}$, а затем их слияние.

\begin{center}
\begin{tabular}{l}
\includegraphics[scale = 0.5, bb= 0 0 180 360]{green_merge.png}   %
\end{tabular}

Рис. 3. Порядок элементов множества $M$ после сортировки в
четверках
\end{center}

Обозначим результаты сортировки четверок номерами с двумя
штрихами. Частичный порядок на множестве $M$ после выполнения
сортировки четверок показан на рис. 3. Он таков, поскольку любой
элемент первой четверки больше каких-то двух элементов второй
четверки, а предпоследний по величине элемент больше трех
элементов второй четверки. Наоборот, каждый элемент второй
четверки уступает каким-то двум элементам первой.

Таким образом, два максимальных и два минимальных элемента $M$
определены, и слияние сводится к доупорядочению четверки $\{10'',
9'', 8'', 7''\}$. Эта процедура выполняется за 3 сравнения, причем
с целью экономии глубины целесообразно начать со сравнения $9''$ и
$8''$: эти элементы вычислены на глубине 8, тогда как два других
--- на глубине 9.

\section{Cортировка множества $M$. Способ ван Вориса}

Как говорилось выше, схема ван Вориса также начинается с
вычисления элементов $13', 11', 10', 7', 6', 4'$ на глубине 5. Но
поскольку элементы $8', 9'$ на этой глубине еще не готовы, на
следующем уровне выполняются дополнительные сравнения в парах
(13', 6'), (11', 7'), (10', 4'). Результаты сравнений занумеруем
числами с двумя штрихами. Теперь после шести уровней компараторов
элементы множества $M$ упорядочены как показано на рис. 4.

\begin{center}
\begin{tabular}{l}
\includegraphics[scale = 0.5, bb= 0 0 372 186]{level6.png}   %
\end{tabular}

Рис. 4. Порядок элементов множества $M$ после предварительных
сравнений
\end{center}

Легко видеть, что каждый элемент верхней четверки $\{8', 13'',
11'', 10''\}$ больше каких-то трех элементов нижней четверки
$\{4'', 6'', 7'', 9'\}$ и наоборот: каждый элемент нижней четверки
меньше каких-то трех элементов верхней четверки.

Поэтому три максимальных элемента множества $M$ относятся к
верхней четверке, три минимальных --- к нижней, а два средних
являются минимальным элементом верхней четверки и максимальным
элементом нижней.

Схема ван Вориса завершается сортировкой указанных четверок и
одним дополнительным сравнением для определения порядка средних
элементов.

\section{Приложение}

На рис. 5, 6 приведены схемы Грина и ван Вориса. Вертикальная
линия отделяет фазу приближенной сортировки в частичный порядок
булева куба. Цифрой 1 подписан блок сортировки слоя I, цифрой 2
--- блок сортировки слоя III, цифрой 3 --- блок сортировки
четверки $\{13', 11', 10', 8'\}$ или $\{13'', 11'', 10'', 8'\}$,
цифрой 4 --- блок сортировки четверки $\{9', 7', 6', 4'\}$ или
$\{9', 7'', 6'', 4''\}$, цифрой 5 --- блок слияния в схеме Грина.

\begin{center}
\begin{tabular}{l}
\includegraphics[scale = 0.4, bb= 0 0 945 497]{green_sort16.png}   %
\end{tabular}

Рис. 5. Схема Грина
\end{center}

\begin{center}
\begin{tabular}{l}
\includegraphics[scale = 0.4, bb= 0 0 976 504]{vv_sort16.png}   %
\end{tabular}

Рис. 6. Схема ван Вориса
\end{center}

В работах~\cite{jui95,cm02,val13} можно найти примеры построенных
при помощи компьютера схем глубины 10 из 60 компараторов с
несколько отличной структурой. В частности, в них урезана фаза
приближенной сортировки Грина.

Работа выполнена при поддержке РФФИ, проект 17-01-00485а.